# Spin structure and magnetic frustration in multiferroic $R\text{Mn}_2\text{O}_5$ ($R$ = Tb, Ho, Dy)

G.R. Blake,[1,2] L.C. Chapon,[1] P.G. Radaelli,[1,3] S. Park,[4] N. Hur,[4] S-W. Cheong,[4] and J. Rodríguez-Carvajal[5]

[1] ISIS Facility, Rutherford Appleton Laboratory-CCLRC, Chilton, Didcot, Oxfordshire, OX11 0QX, United Kingdom

[2] Materials Science Division, Argonne National Laboratory, Argonne, IL 60439, USA

[3] Department of Physics and Astronomy, University College London, Gower Street, London WC1E 6BT, United Kingdom

[4] Department of Physics and Astronomy, Rutgers University, Piscataway, New Jersey 08854, USA

[5] Laboratoire Léon Brillouin (CEA-CNRS), CEA/Saclay, 91191 Gif-sur-Yvette Cedex, France

## ABSTRACT

We have studied the crystal and magnetic structures of the magnetoelectric materials $R\text{Mn}_2\text{O}_5$ ($R$ = Tb, Ho, Dy) using neutron diffraction as a function of temperature. All three materials display incommensurate antiferromagnetic ordering below 40 K, becoming commensurate on further cooling. For $R$ = Tb, Ho, a commensurate-incommensurate transition takes place at low temperatures. The commensurate magnetic structures have been solved and are discussed in terms of competing exchange interactions. The spin configuration within the $ab$ plane is essentially the same for each system, and the radius of $R$ determines the sign of the magnetic exchange between adjacent planes. The inherent magnetic frustration in these materials is lifted by a small lattice distortion, primarily involving shifts of the $\text{Mn}^{3+}$ cations and giving rise to a canted antiferroelectric phase.

## I. INTRODUCTION

There is a great deal of current interest in materials that exhibit interplay between lattice distortions and electrical and magnetic ordering. In particular, the small group of materials known as magnetoelectrics, in which magnetism and ferroelectricity coexist and are mutually coupled, are being extensively investigated. These materials display phenomena such as the control of electrical polarization by the application of an external magnetic field, providing an additional degree of freedom for the design of new devices. Such behavior has recently been found in $TbMnO_3$,[1] $DyMnO_3$[2] and $TbMn_2O_5$;[3] these systems all have incommensurate magnetic order caused by competing magnetic exchange interactions, which increasingly appears to be a feature that can give rise to magnetoelectric properties.

The manganese oxides with general formula $RMn_2O_5$ ($R$ = La, Y, Bi or rare-earth) are insulators and consist of linked $Mn^{4+}O_6$ octahedra and $Mn^{3+}O_5$ pyramids (Fig. 1), adopting space group *Pbam*. The octahedra share edges to form ribbons parallel to the *c* axis, adjacent ribbons being linked by pairs of corner-sharing pyramids. These materials have been studied since the 1960s due to their complex magnetic structures, but more recently have been found to exhibit spontaneous electrical polarization, the onset of which occurs just below the antiferromagnetic (AFM) ordering temperature ($T_N$).[4-7] Although the magnitude of this polarization (**P**) is two or three orders of magnitude smaller than in typical ferroelectrics, there is growing evidence that the polarization is strongly coupled to the magnetic order. Recent studies of $RMn_2O_5$ materials have revealed remarkable magnetoelectric properties. In $TbMn_2O_5$ the direction of **P** can be repeatedly reversed at 3 K, without any loss in magnitude, by the periodic variation of an external magnetic field between 0 and 2 T.[3] The application of a magnetic field also enhances the dielectric constant $(\varepsilon)$ of $RMn_2O_5$ ($R$ = Tb, Ho, Dy), by as much as 109% in the case of $DyMn_2O_5$.[8] On cooling these materials below $T_N$, multiple phase transitions involving changes in the magnetic propagation vector $\mathbf{k} = (k_x, 0, k_z)$, where $k_x \approx 1/2$, are a common feature and often coincide with pronounced anomalies in $\varepsilon$, **P**, bulk magnetization and specific heat.[5-7] However, the precise nature of the interplay and coupling between the crystal structure, ferroelectricity and magnetic ordering remains rather unclear. One would expect a phase transition to a structure with a non-centrosymmetric space group to occur at the onset of ferroelectricity, but studies thus far have failed to find direct evidence of such changes.[7,9] Furthermore, due to their complexity, detailed determinations of the $RMn_2O_5$ magnetic structures are lacking for all but the simplest cases. In order to gain a better understanding of these complex systems, we recently carried out a neutron diffraction study of $TbMn_2O_5$ which revealed unambiguous correlations between anomalies in ε and changes in periodicity of the spin structure on varying the temperature.[10] The Mn spins and a small proportion of the Tb spins order at $T_N$ = 43 K, slightly above the ferroelectric ordering temperature at $T_C$ = 38 K. Our data showed that the magnetic structure is incommensurate ($\mathbf{k}$ = (~0.50,0,0.30)) immediately below $T_N$, becoming commensurate with $\mathbf{k}$ = (1/2,0,1/4) on cooling through a "lock-in" temperature of 33 K. Unusually, a commensurate to incommensurate ($\mathbf{k}$ = (0.48,0,0.32)) transition takes place at 24 K, at which temperature a large jump in ε and a rapid decrease in **P** have been observed.[3] Ordering of the remaining Tb spins then takes place on cooling below 9 K, coinciding with a recovery in **P**. $TbMn_2O_5$ is a geometrically frustrated system, in which the favorable magnetic exchange interactions cannot all be satisfied simultaneously. In this scenario, small displacements of the $Mn^{3+}$ cations would lift the corresponding magnetic degeneracy and reduce the exchange energy. The unusually small value of **P** would then result from a "canted antiferroelectric" arrangement of atomic displacement vectors. In order to confirm the above hypothesis, it is important to study other members of the $R$Mn$_2$O$_5$ series and to investigate the role played by the rare-earth cation; the magnetic propagation vector in this series of materials depends strongly on $R$ as well as on temperature. Here we present further details of magnetically frustrated TbMn$_2$O$_5$ and report on the magnetic and crystal structures of HoMn$_2$O$_5$ and DyMn$_2$O$_5$. We show that the Mn spins order in essentially the same configuration within the *ab* plane regardless of $R$, and that the magnetic structure is consistent with the lowering of symmetry in the ferroelectric phase from *Pbam* to *Pb2$_1$m* that has been predicted[9] by group theoretical analysis. We also present energy calculations of the collinear magnetic ground state; these indicate that the observed magnetic structure cannot be stabilized in the *Pbam* space group.

## II. EXPERIMENT

Polycrystalline $R$Mn$_2$O$_5$ was prepared through conventional solid-state reaction in an oxygen environment. Samples were sintered at 1120 °C for 40 hours with intermediate grindings, followed by slow cooling. Neutron powder diffraction data were collected using the GEM diffractometer at the ISIS facility. A helium cryostat was employed to vary the temperature between 2 K and 300 K. Determinations of the nuclear and magnetic structures were carried out using the GSAS and FULLPROF programs, respectively.[11]

## III. NUCLEAR STRUCTURES

Refinements of the nuclear structures of all three $R$Mn$_2$O$_5$ materials were carried out in the centrosymmetric space group *Pbam*. The ferroelectric transition temperatures are 38 K for Tb,[3] 40 K for Ho and 39 K for Dy.[8] Although the structures must be non-centrosymmetric in the ferroelectric phase, our data do not provide direct evidence for the lowering of symmetry. We did not observe any nuclear superlattice peaks that would indicate a modulation of the ferroelectric phase. This is consistent with a previous structural study of ferroelectric YMn$_2$O$_5$[9] using synchrotron X-ray diffraction on single crystals, which failed to find evidence for the expected symmetry lowering. However, we did observe anomalies in the lattice parameters (Tb and Dy) and ADPs (Tb). Further details of the TbMn$_2$O$_5$ refinements are presented in Ref. 10. The temperature dependence of the lattice parameters for the Ho and Dy compounds are shown in Figs. 2(a) and 2(b). The trends for HoMn$_2$O$_5$ are broadly similar to those for the Tb sample, with the *a* and *c* parameters becoming essentially constant below 30 K. However, unlike in the case of Tb there is no sign of any anomaly in the *b* parameter. DyMn$_2$O$_5$ shows a much larger structural response than the other two samples. A sharp reversal of the slope of the *c* lattice parameter occurs on cooling below 25 K, which results in a slight negative thermal expansion of the unit cell as a whole. This coincides with anomalies in the specific heat and dielectric constant.[8] In addition, the *a* parameter appears to have a small anomaly at ~15 K, but higher resolution data are clearly needed to confirm if this feature is significant. The cause of these features remains unclear at present- it is possible that a modulation of the lattice could occur in the low-temperature phase, but with our current data we are unable to speculate further. For both the Ho and Dy samples, good fits were obtained in space group *Pbam* at all temperatures measured, and no anomalies in the ADPs were apparent. Any low-temperature deviation from *Pbam* symmetry is thus very small in both cases. This is consistent with the extremely weak nature of the polarization, two or three orders of magnitude smaller than in typical ferroelectrics, which is not expected to give large atomic displacements. The high-temperature structures of the Tb and Ho samples agree well with those reported by Alonso et al.,[12] and the Dy structure is essentially identical. Table I lists some bond distances and angles at 60 K that are relevant to the discussion of the magnetic structures below.

**TABLE I: Selected bond distances (Å) and angles (degrees) at 60 K.**

| Distance/angle | TbMn$_2$O$_5$ | HoMn$_2$O$_5$ | DyMn$_2$O$_5$ |
|---|---|---|---|
| Mn$^{4+}$-O2 | 1.931(2) | 1.926(2) | 1.922(4) |
| Mn$^{4+}$-O3 | 1.861(2) | 1.865(2) | 1.873(4) |
| Mn$^{4+}$-O4 | 1.911(1) | 1.907(1) | 1.919(2) |
| Mn$^{3+}$-O1 | 1.917(2) | 1.919(2) | 1.920(4) |
| Mn$^{3+}$-O3 | 2.027(2) | 2.012(3) | 2.018(5) |
| Mn$^{3+}$-O4 | 1.903(2) | 1.903(2) | 1.896(4) |
| Mn$^{4+}$-Mn$^{4+}$ (at "Mn$^{3+}$ layer") | 2.760(4) | 2.777(5) | 2.788(10) |
| Mn$^{4+}$-Mn$^{4+}$ (at "$R$ layer") | 2.902(4) | 2.887(5) | 2.879(10) |
| Mn$^{3+}$-Mn$^{3+}$ | 2.842(4) | 2.830(5) | 2.846(9) |
| Mn$^{4+}$-O2- Mn$^{4+}$ (J1) | 97.45(11) | 97.10(13) | 97.0(3) |
| Mn$^{4+}$-O3- Mn$^{4+}$ (J2) | 95.72(11) | 96.24(14) | 96.2(3) |
| Mn$^{4+}$-O4- Mn$^{3+}$ (J3) | 123.09(9) | 122.59(11) | 122.6(2) |
| Mn$^{4+}$-O3- Mn$^{3+}$ (J4) | 131.68(6) | 131.32(7) | 131.4(1) |
| Mn$^{3+}$-O1- Mn$^{3+}$ (J5) | 95.70(10) | 95.03(12) | 95.7(2) |

## IV. MAGNETIC STRUCTURES

The magnetic propagation vector for HoMn$_2$O$_5$ is plotted as a function of temperature in Fig. 3(a); the trend is rather similar to that for TbMn$_2$O$_5$.[10] Immediately below $T_N$ = 44 K, the magnetic structure is incommensurate and all the magnetic Bragg peaks can be indexed using **k** = (0.480,0,0.245). A transition to a commensurate magnetic structure with **k** = (1/2,0,1/4) then takes place at ~38 K, coinciding with $T_C$, before it becomes incommensurate once again below 18 K with **k** = (0.480,0,0.280), coinciding with anomalies in the specific heat and dielectric constant.[8] Similar sequences of transitions have also been observed in ErMn$_2$O$_5$[6] and YMn$_2$O$_5$.[7] The background, integrated over the Q-range 0.74 to 1.05 Å$^{-1}$, and the integrated intensity of the (110)-**k** magnetic Bragg peak are plotted in Fig. 3(b). The latter curve becomes slightly steeper below 18 K, suggesting that the commensurate-incommensurate transition involves the onset of ordering in the Ho$^{3+}$ sublattice. However, this ordering is likely to be gradual in nature, as the background decreases in essentially linear fashion over the whole temperature range.

For DyMn$_2$O$_5$ the behavior of **k** is rather different to that for $R$ = Tb and Ho. In our neutron diffraction data, magnetic peaks are first apparent above the high background (due to the large incoherent neutron cross-section of Dy) at 32 K, although the true ordering temperature may well be higher.[8] The magnetic structure is incommensurate below 32 K, with **k** = (0.490,0,0.250). The value of **k** remains unchanged on cooling to 8 K, where a transition to a commensurate structure with **k** = (0.5,0,0) takes place.

Although no low-temperature re-entrant incommensurate phase was observed, weak peaks that could be indexed with the propagation vector of the "high-temperature" magnetic phase persist down to 2 K. These either indicate an additional modulation of the "average" magnetic structure, as reported by Wilkinson et al.,[13] or an incomplete phase transition to the low-temperature, commensurate phase. Unfortunately the extra peaks in our data are too weak to allow us to distinguish between the two possible scenarios. Any magnetic contribution to the background is overshadowed by incoherent scattering from Dy, but from the integrated intensity of the (100)+**k** magnetic Bragg peak, shown in Fig. 4, it appears that the degree of order on the rare-earth sublattice increases below the 8 K transition. We note that no magnetic transition is apparent in the vicinity of the specific heat and dielectric constant anomalies at ~25 K.[8]

The magnetic structures of the commensurate phases of all three samples were solved with the help of the simulated annealing method incorporated in FULLPROF, assuming space group *Pbam*. Symmetry analyses were first carried out and are described in the Appendix. However, this revealed that the crystal symmetry imposes very few constraints on the variables to be determined. Specifically, for all three materials pairs of $Mn^{3+}$ atoms (4h) and *R* atoms (4g) at (*x,y,z*) and (*-x,-y,z*) are related such that individual components of the magnetic moments ($m_x$, $m_y$ and $m_z$) can be coupled in either FM or AFM fashion. For $DyMn_2O_5$ only, pairs of $Mn^{4+}$ moments at (0,0.5,*z*), (0,0.5,*-z*) and at (0.5,0,*z*), (0.5,0,*-z*) are related such that both $m_x$ and $m_y$ are coupled in either AFM or FM fashion; $m_z$ is then coupled in the opposite fashion.

Thus, symmetry analysis does little to reduce the number of independent variables in the problem (in fact, the true symmetry in the magnetically ordered regime is expected to be lower than *Pbam*). It was therefore necessary to introduce additional constraints in the simulated annealing procedure. Firstly, the magnitudes of the magnetic moments $|m_{total}|$ were constrained to be equal for all atoms of the same type ($Mn^{3+}$, $Mn^{4+}$, *R*). Secondly, the phases of the spin density waves (SDWs) for $TbMn_2O_5$ and $HoMn_2O_5$ were constrained to be the same for all moments associated with a given crystallographic site. Starting configurations containing the possible linear combinations of AFM or FM-coupled $m_x$, $m_y$ and $m_z$ components were then formulated (limited only by the small number of symmetry constraints described above), and an input file for each was written. The experimental data used in the simulated annealing runs consisted of a list of integrated intensities of purely magnetic peaks extracted from the powder patterns by full-profile fitting (between 50 and 100 reflections). It soon became apparent from preliminary simulated annealing runs that all moments lie in the *ab* plane for all three materials, thus simplifying the problem. The models giving the best fits to the integrated intensity data were selected for Rietveld refinement using FULLPROF. For the Tb and Ho compounds the $Mn^{3+}$ and $Mn^{4+}$ SDW phases obtained from simulated annealing were essentially equal, and that of *R* was shifted by almost exactly $\pi/4$; all phases were subsequently fixed in the refinements to rational fractions of $\pi$. In all three cases, stable refinements were only obtained when all $|m_{total}|$ for Mn cations of the same charge were constrained to be equal.

The best models obtained from the Rietveld refinements were very similar for $TbMn_2O_5$ and $HoMn_2O_5$. For $DyMn_2O_5$ the best solution was essentially the same as that reported by Wilkinson et al.[13] The refined magnetic parameters are summarized in Tables II(a) to (c) and schematic representations of the magnetic structures in the *ab* plane are shown in Fig. 5. The observed, calculated and difference neutron diffraction profiles are shown in Fig. 6.

**TABLE II(a): TbMn$_2$O$_5$ magnetic structure at 27 K; propagation vector $\underline{k}$ = (0.5,0,0.25), all moments are in *ab* plane.**

| Atom | x | y | z | Moment (μ$_B$) | Phi (degrees) | Phase (2π) |
|---|---|---|---|---|---|---|
| Mn$^{4+}$ (1) | 0 | 0.5 | 0.2557 | 1.86(7) | 163(7) | 0.125 |
| Mn$^{4+}$ (2) | 0 | 0.5 | 0.7443 | 1.86(7) | 163(7) | 0.125 |
| Mn$^{4+}$ (3) | 0.5 | 0 | 0.2557 | 1.86(7) | 160(6) | 0.125 |
| Mn$^{4+}$ (4) | 0.5 | 0 | 0.7443 | 1.86(7) | 160(6) | 0.125 |
| Mn$^{3+}$ (1) | 0.0886 | 0.8505 | 0.5 | 2.41(5) | 354(8) | 0.125 |
| Mn$^{3+}$ (2) | 0.4114 | 0.3505 | 0.5 | 2.41(5) | 329(8) | 0.125 |
| Mn$^{3+}$ (3) | 0.5886 | 0.6495 | 0.5 | 2.41(5) | 149(8) | 0.125 |
| Mn$^{3+}$ (4) | 0.9114 | 0.1495 | 0.5 | 2.41(5) | 354(8) | 0.125 |
| Tb$^{3+}$ (1) | 0.1396 | 0.1719 | 0 | 1.18(9) | 349(18) | 0 |
| Tb$^{3+}$ (2) | 0.3604 | 0.6719 | 0 | 2.24(7) | 338(8) | 0 |
| Tb$^{3+}$ (3) | 0.6396 | 0.3281 | 0 | 2.24(7) | 338(8) | 0 |
| Tb$^{3+}$ (4) | 0.8604 | 0.8281 | 0 | 1.18(9) | 349(18) | 0 |

**TABLE II(b): HoMn$_2$O$_5$ magnetic structure at 26 K; propagation vector $\underline{k}$ = (0.5,0,0.25), all moments are in *ab* plane.**

| Atom | x | y | z | Moment (μ$_B$) | Phi (degrees) | Phase (2π) |
|---|---|---|---|---|---|---|
| Mn$^{4+}$ (1) | 0 | 0.5 | 0.2558 | 2.20(9) | 169(14) | 0.125 |
| Mn$^{4+}$ (2) | 0 | 0.5 | 0.7442 | 2.20(9) | 169(14) | 0.125 |
| Mn$^{4+}$ (3) | 0.5 | 0 | 0.2558 | 2.20(9) | 162(5) | 0.125 |
| Mn$^{4+}$ (4) | 0.5 | 0 | 0.7442 | 2.20(9) | 162(5) | 0.125 |
| Mn$^{3+}$ (1) | 0.0885 | 0.8490 | 0.5 | 2.53(9) | 3(6) | 0.125 |
| Mn$^{3+}$ (2) | 0.4115 | 0.3490 | 0.5 | 2.53(9) | 344(15) | 0.125 |
| Mn$^{3+}$ (3) | 0.5885 | 0.6510 | 0.5 | 2.53(9) | 164(15) | 0.125 |
| Mn$^{3+}$ (4) | 0.9115 | 0.1510 | 0.5 | 2.53(9) | 3(6) | 0.125 |
| Ho$^{3+}$ (1) | 0.1392 | 0.1713 | 0 | 1.86(29) | 307(8) | 0 |
| Ho$^{3+}$ (2) | 0.3608 | 0.6713 | 0 | 1.32(23) | 28(11) | 0 |
| Ho$^{3+}$ (3) | 0.6392 | 0.3287 | 0 | 1.32(23) | 28(11) | 0 |
| Ho$^{3+}$ (4) | 0.8606 | 0.8287 | 0 | 1.86(29) | 307(8) | 0 |

**TABLE II(c): DyMn$_2$O$_5$ magnetic structure at 2 K; propagation vector $\underline{k}$ = (0.5,0,0), all moments are in *ab* plane**

| Atom | x | y | z | Moment ($\mu_B$) | Phi (degrees) |
|---|---|---|---|---|---|
| Mn$^{4+}$ (1) | 0 | 0.5 | 0.2521 | 1.27(15) | 299(9) |
| Mn$^{4+}$ (2) | 0 | 0.5 | 0.7479 | 1.27(15) | 299(9) |
| Mn$^{4+}$ (3) | 0.5 | 0 | 0.2521 | 1.27(15) | 61(9) |
| Mn$^{4+}$ (4) | 0.5 | 0 | 0.7479 | 1.27(15) | 61(9) |
| Mn$^{3+}$ (1) | 0.0759 | 0.8447 | 0.5 | 1.7(4) | 244(30) |
| Mn$^{3+}$ (2) | 0.4241 | 0.3447 | 0.5 | 1.7(4) | 116(30) |
| Mn$^{3+}$ (3) | 0.5759 | 0.6553 | 0.5 | 1.7(4) | 296(30) |
| Mn$^{3+}$ (4) | 0.9241 | 0.1553 | 0.5 | 1.7(4) | 244(30) |
| Dy$^{3+}$ (1) | 0.1389 | 0.1729 | 0 | 5.68(13) | 270.0 |
| Dy$^{3+}$ (2) | 0.3611 | 0.6729 | 0 | 5.68(13) | 90.0 |
| Dy$^{3+}$ (3) | 0.6389 | 0.3271 | 0 | 5.68(13) | 270.0 |
| Dy$^{3+}$ (4) | 0.8611 | 0.8271 | 0 | 5.68(13) | 270.0 |

The configurations of the ordered Mn moments in all three samples are consistent with the prediction by Kagomiya et al.[9] of a lowering of the crystal symmetry at least down to to $Pb2_1m$ in the ferroelectric phase, based on a group theoretical analysis of possible Mn$^{3+}$ displacements that could give rise to polarization along the *b* axis. This is best shown by constructing a "toy model" of the magnetic structure with exact magnetic space group symmetry $P_{2a}b'2_1m'$ (using the Shubnikov formalism, see Appendix). This model provides a good description of the configuration of the Mn spins in the case of DyMn$_2$O$_5$. If the *total* symmetry (magnetic +crystal) of the system is $P_{2a}b'2_1m'$, the symmetry of the *crystal* structure is $Pb2_1m$, which is the corresponding paramagnetic supergroup. This toy model provides a link between the magnetic structure and the proposed lowering of symmetry in the ferroelectric phase, but is clearly an oversimplification: in reality, there is an additional modulation along the *c* axis and "misalignment" of the Mn spins by up to 30° in TbMn$_2$O$_5$ and HoMn$_2$O$_5$. This suggests that the real crystal symmetry may be even lower than $Pb2_1m$. Further details are given in the Appendix.

## V. MAGNETIC EXCHANGE INTERACTIONS

The spins lie in the *ab* plane for all three materials. Within the *ab* plane, it can be seen in Fig. 5 that two zig-zag chains per unit cell of AFM-coupled nearest-neighbor Mn$^{4+}$ and Mn$^{3+}$ run in a direction parallel to the *a* axis The canting angles of the AFM-coupled spins in these chains are essentially zero within error bars, being refined as 14(11)° and 14(10)° for Tb, 5(21)° and 21(8)° for Ho, and 3(31)° and 3(31)° for Dy. In all three materials the magnetic moments of both Mn$^{4+}$ and Mn$^{3+}$ are much lower than expected, suggesting that a degree of frustration is present. This is unsurprising given the nature of the lattice geometry, which gives rise to competition between different magnetic exchange interactions; five nearest-neighbor interactions can be identified, shown in Fig. 1. The Mn-O-Mn bond angles associated with these interactions in the case of superexchange via an oxygen atom are listed in Table II. Looking at the exchange interactions relevant to the *ab* plane, J3 and J4 are associated with bond angles that are close to the crossover point between AFM and FM superexchange interactions (~123° and ~131°, respectively), according to the Goodenough-Kanamori-Anderson (GKA) rules.[14] It appears that |J4| > |J3| and that J4 is always AFM, giving rise to the zig-zag chains in which pairs of J4 interactions are separated by an AFM J5 interaction. This results in the ubiquitous doubling of the *a* axis in these materials. However, the frustrated topology makes it impossible to satisfy all of the favorable interactions simultaneously, and every $Mn^{4+}$ moment has one nearest neighbor $Mn^{3+}$ moment in the *b* direction with the "wrong" sign. Competition between different exchange interactions is not confined to the *ab* plane; J3/J4 will favour a FM alignment of $Mn^{4+}$ spins in adjacent edge-shared octahedra, while the weak superexchange associated with J2 is expected to support an AFM arrangement. In all three of our materials |J3| > |J2| and |J4| > |J2|, thus the alignment is always FM. The magnetic structures of the $RMn_2O_5$ series mainly differ in their periodicity along *c*, which is most likely determined by the radius of *R*. Although the arrangement of Mn spins within the *ab* plane is essentially insensitive to *R*, the radius of *R* determines the nature of J1, the interaction between adjacent $Mn^{4+}$ spins in edge-shared octahedra at the "*R* layer". Competition is expected here between weak superexchange (involving a $Mn^{4+}$-O-$Mn^{4+}$ bond angle of ~97°) and direct exchange. The $Mn^{4+}$-O1-$Mn^{4+}$ bond angles become smaller as the size of *R* decreases; there is a ~0.5° difference between Tb and Dy (Table I). The $Mn^{4+}$-$Mn^{4+}$ distances also decrease by ~0.02 Å from Tb to Dy, most likely strengthening the direct exchange interaction. Each $Mn^{4+}$ here is linked to a pair of $Mn^{3+}$ cations through J3 and J4, and so these two interactions may also play a role in the spin configuration at the "*R* layer" and hence in the final value of $k_z$. The competing interactions combine such that adjacent $Mn^{4+}$ spins either side of the "*R* layer" are FM for Dy, retaining the original lattice periodicity along *c* ($k_z = 0$), AFM for the larger $Bi^{3+}$ cation, giving a two-fold superstructure ($k_z = 0.5$),[15] and alternately FM and AFM for Tb, Ho, Y[7,16] and Er,[6] resulting in a four-fold superstructure ($k_z = 0.25$). In the case of the commensurate Tb and Ho phases, partial ordering of the rare-earth sublattice appears to be induced by the ordered Mn sublattice, and is influenced in particular by the signs of the $Mn^{4+}$ moments either side of the "*R* layer". A non-zero moment on Tb or Ho only occurs when adjacent $Mn^{4+}$ spins are FM; the alternating FM and AFM linkages result in a zero moment on every second layer of Tb and Ho atoms and in a phase shift of $\pi/4$ for the Tb/Ho SDW with respect to that of both Mn sites. One would also expect the alternating nature of these $Mn^{4+}$-$Mn^{4+}$ linkages to cause a small positional modulation of *R* and O2, evidence for which was found in the $TbMn_2O_5$ ADPs.[10] A weak modulation of bond lengths would thus tend to stabilize the four-fold magnetic superstructure along *c*. In the Dy sample, adjacent $Mn^{4+}$ spins are always FM and no modulation of bond lengths along *c* is necessary to stabilize the magnetic structure in this direction.

To generalize further, the particular topology of the magnetic Mn sublattices in $RMn_2O_5$ is the source of a complex interplay of exchange interactions. The most important closed loops (circuits), constructed using the Mn atoms as nodes, have an odd number of nodes. With negative exchange interactions these odd circuits give rise to frustration. Alternatively, the magnetic structure of the $RMn_2O_5$ materials in the *ab* plane can be visualized in terms of an AFM square lattice of $Mn^{4+}$ with asymmetric next-nearest-neighbour (NNN) interactions, a simple geometrically frustrated system (Fig. 7). A hierarchy of three NNN interactions can be identified:
Interaction1 < 0, interaction2 > 0, interaction3 > 0, and |interaction3| > |interaction2|.

The NNN interaction along the *a* axis is thus stronger than that along the *b* axis, and the zig-zag AFM chains parallel to *a* are always stabilized.

In such frustrated systems a structural distortion will tend to occur in order to give a non-degenerate ground state. Here the frustration appears to be responsible for inducing the transition to the ferroelectric phases. Although we have no direct crystallographic evidence for a lowering of the symmetry, the ADP anomalies observed in TbMn$_2$O$_5$ close to the ferroelectric ordering temperature suggest that coordinated shifts of the Mn$^{3+}$ cations take place to give a canted antiferroelectric structure and a net polarization along the *b* axis.[10] A structural transition to the space group *Pb2$_1$m*, as previously predicted using group-theoretical considerations,[9] is consistent with this scenario. The Mn$^{3+}$ site would be split into two inequivalent sites, inducing a modulation in the Mn$^{4+}$-O-Mn$^{3+}$ bond angles in order to strengthen exchange interactions with the "right" sign and weaken those with the "wrong" sign. The same scenario is almost certainly valid for the Ho and Dy compounds, but the extremely small structural distortions involved, suggested by the small magnitude of **P**, are on the limit of detection using conventional diffraction methods.

The nature of the incommensurate magnetic phases remains rather unclear. Here each of the 8 Mn atoms and 4 *R* atoms in the crystallographic unit cell is allowed to have its own spin amplitude and phase, and there are no obvious phase relations between the SDWs of different atoms. We were unable to obtain unique solutions for the incommensurate magnetic structures and will probably require single crystal data in order to attack this problem in a systematic manner. We speculate that the incommensurate phases might result from reversal of the AFM zig-zag chains along the *a* axis; one or both chains might be reversed, giving rise to 4 possible magnetic configurations per Mn$^{3+}$/Mn$^{4+}$ layer. The incommensurate phases might then contain variable mixtures of the different configurations. If one of the two chains is reversed, **P** would also be reversed from the (b+) to (b-) direction, giving a possible explanation for the observed temperature dependence of **P** in these materials.[8]

## VI. MAGNETIC PHASE DIAGRAM

We attempted to clarify the relationship between the strengths of the various Mn-Mn exchange interactions by calculating the ground-state collinear magnetic configuration for a given set of isotropic exchange interactions. This calculation was performed using the program ENERMAG.[17] The energy of the ground-state configuration is given by the lowest eigenvalue, $\lambda(\mathbf{k},\{J_{ij}\})$, of the Fourier transform of the exchange integral matrix, $\xi(\mathbf{k},\{J_{ij}\})$, where $\{J_{ij}\}$ is the set of exchange integrals. Thus, $\lambda(\mathbf{k},\{J_{ij}\})$ is minimized with respect to **k**, which is then the propagation vector of the ground-state configuration; for commensurate structures the sequence of signs of the corresponding eigenvector components gives the spin configuration.

Super-exchange and super-super-exchange pathways were first calculated using the program SIMBO.[17] We used both the atomic positions of TbMn$_2$O$_5$ refined in *Pbam* and in the predicted space group *Pb2$_1$m* as input. The output from SIMBO was then used as the input for ENERMAG. We decided to focus only on the spin configuration within the *ab* plane, since the configuration parallel to *c*, involving interactions J1 and J2 (Fig. 1), is determined only by the radius of the rare-earth cation; calculations were therefore carried out setting J1 = J2 = 0 in the ENERMAG input file. For the same reason, the propagation vector component $k_z$ was set to zero, while $k_x$ and $k_y$ were allowed to vary in the range 0 to 0.5 during the minimization process for each set of $\{J_{ij}\}$. Propagation vectors will thus be referred to in the discussion below as $\mathbf{k} = (k_x, k_y)$.

*Pbam calculations*

Calculations were carried out by setting J4 (shown in Fig. 1) to either a positive or negative value, then systematically varying J3 and J5 in units of J4. Selected parts of the resulting phase diagram are shown schematically in Fig. 8. Various different magnetic structures are predicted, and the spin configurations in regions of the phase diagram where there is no magnetic degeneracy are listed in Table III. There are four different $\mathbf{k} = (0,0)$ structures predicted at positive (and sometimes low negative) values of J5, a range of degenerate $\mathbf{k} = (0.5, 0.5)$ structures at large negative values of J5, and various incommensurate structures where $k_x$, $k_y$ or both components deviate from 0 or 0.5. The commensurate $\mathbf{k} = (0.5, 0)$ structure experimentally observed in $R$Mn$_2$O$_5$ ($R$ = Tb, Ho, Dy) corresponds to configuration 3a in Table III, which is only realized in the plane of the three-dimensional phase diagram formed by the exchange interactions J3 = 0, J4 < 0 and J5 < 0 (represented by the vertical dotted line in the J4 < 0 diagrams). Even here, the structure appears to be rather poorly defined in the $b$ direction, since although $k_y = 0$ on average, the value fluctuates significantly from point to point in the plane. Indeed, it is difficult to envisage how long-range order along the $b$ axis could occur when J3 = 0. These observations strengthen our qualitative observation that the experimentally observed structure cannot be stabilized in *Pbam* symmetry, due to the presence of frustration.

**TABLE III: Predicted collinear $R$Mn$_2$O$_5$ magnetic configurations; atom numbers refer to those in Tables II(a) to II(c), and "+" and "-" represent the direction of spins.**

| Propagation vector: | (0,0) | | | | | | (0.5,0) | | |
|---|---|---|---|---|---|---|---|---|---|
| Configuration: | 1a | 1b | 1c | 1d | 1e | 1f | **3a** | 3b | 3c |
| Mn$^{4+}$ (1) | + | + | + | + | + | + | + | + | + |
| Mn$^{4+}$ (2) | + | + | + | + | + | + | + | + | + |
| Mn$^{4+}$ (3) | + | - | - | + | + | - | - | - | + |
| Mn$^{4+}$ (4) | + | - | - | + | + | - | - | - | + |
| Mn$^{3+}$ (1) | - | + | - | + | - | + | + | - | - |
| Mn$^{3+}$ (2) | - | - | + | + | - | - | - | + | - |
| Mn$^{3+}$ (3) | - | - | + | + | + | + | + | - | - |
| Mn$^{3+}$ (4) | - | + | - | + | + | - | + | - | + |

*Pb2$_1$m calculations*

The proposed structural distortion giving rise to $Pb2_1m$ symmetry[9] would split both J3 and J4 into two inequivalent interactions. The split J3 interactions will hereafter be referred to as J3a and J3b, and the split J4 interactions as J4a and J4b. Neglecting J1 and J2, which were again fixed to zero, there are 5 variable interactions in $Pb2_1m$ symmetry. The calculation time for a full five-dimensional phase diagram would be prohibitive, and calculations were thus carried out by constraining either J3a = J3b or J4a = J4b and fixing the value of the constrained pair, then varying the other three parameters. A selection of the schematic phase diagrams obtained is shown in Fig. 9: J3a, J3b and J5 were varied in set (a) and J4a, J4b and J5 were varied in set (b). The phase diagrams are plotted in units of the pair of fixed interactions.

The phase diagrams are generally more complicated than in the *Pbam* case. The most important result is that splitting either the *Pbam* J3 or J4 interaction stabilizes regions of the phase diagram containing the experimentally observed **k** = (0.5,0) magnetic configuration (3a). The splitting in energy between J3a and J3b, or between J4a and J4b, that is required to stabilize the "3a phase" becomes smaller as the J4a/J3a ratio increases. Since the structural distortion from *Pbam* symmetry is very small, the differences in energy between J3a and J3b, and between J4a and J4b, are also likely to be small. It thus appears that J3a and J3b are weak interactions in comparison to J4a and J4b, as proposed in the discussion of the magnetic structure above. In Fig. 9(a) it may be seen that if J3a and J3b have the same magnitude but opposite sign, the experimentally observed 3a configuration is always stable for J5 < 0. This scenario could arise if a small structural distortion causes a modulation in the $Mn^{4+}$-O4-$Mn^{3+}$ bond angle; this angle is close to the AFM-FM crossover point, and a distortion to $Pb2_1m$ symmetry could lead to J3a and J3b having opposite signs.

A feature common to many of the phase diagram "slices" in Fig. 9 is the existence of two distinct areas of configuration 3a at negative values of J5 that are separated by an incommensurate (IC) region. The IC region becomes "narrower" in energy as the J4a/J3a ratio increases, that is, as J3a becomes weaker. This particular IC region has **k** = ($k_x$,0), 0 < $k_x$ < 0.5, and may correspond to the ($k_x$,0,$k_z$) phases reported for many of the $R$Mn$_2$O$_5$ materials.

## VII. SUMMARY

The magnetoelectric materials $R$Mn$_2$O$_5$ (R = Tb, Ho, Dy) all display multiple magnetic phase transitions. A variety of magnetic ground states, both commensurate and incommensurate, appear to lie very close to each other in energy, giving complex phase relations. However, the spin configuration within the *ab* plane of the commensurate phases is essentially the same for each system; the radius of *R* determines the sign of the magnetic exchange between adjacent planes. The inherent magnetic frustration caused by the lattice geometry is lifted by small shifts of the $Mn^{3+}$ cations. Both the magnetic structures and our energy calculations suggest that the space group symmetry is most likely lowered from *Pbam* to $Pb2_1m$ and that a canted antiferroelectric state is induced with a small net polarization parallel to the *b* axis.

## ACKNOWLEDGMENT

This work was sponsored in part by the U.S. Department of Energy Office of Science under Contract No. W-31-109-ENG-38 and by the NSF-DMR-0405682.

# APPENDIX: SYMMETRY ANALYSIS OF THE MAGNETIC STRUCTURE OF $R$MN$_2$O$_5$.

The propagation vector of the magnetic structure for all samples investigated here is $\mathbf{k} = (1/2, 0, k_z)$, labelled $\{k_{16}\}$ in Kovalev's notation. Four rotational elements of the space group *Pbam* leave this propagation vector invariant: $\{1|000\}$, $\{2_z|000\}$, $\{m_x|000\}$ and $\{m_y|000\}$, using the notation of the International Tables. The single irreducible representation of the group of the propagation vector $G_k$ is shown in Table IV where the symmetry elements are labelled according to the setting of the International Tables.

**Table IV**

| Symmetry elements of $G_k$ | $\{1\|000\}$ | $\{2_z\|000\}$ | $\{m_x\|0½0\}$ | $\{m_y\|½00\}$ |
|---|---|---|---|---|
| $\Gamma_1$ | $\begin{pmatrix}1 & 0\\0 & 1\end{pmatrix}$ | $\begin{pmatrix}1 & 0\\0 & -1\end{pmatrix}$ | $\begin{pmatrix}0 & i\\-i & 0\end{pmatrix}$ | $\begin{pmatrix}0 & -i\\-i & 0\end{pmatrix}$ |
| $\Gamma_1$ (real matrices) | $\begin{pmatrix}1 & 0\\0 & 1\end{pmatrix}$ | $\begin{pmatrix}1 & 0\\0 & -1\end{pmatrix}$ | $\begin{pmatrix}0 & 1\\1 & 0\end{pmatrix}$ | $\begin{pmatrix}0 & -1\\1 & 0\end{pmatrix}$ |
| $\chi(\Gamma_1)$ | 2 | 0 | 0 | 0 |
| **Atom in site 4(f)-Orbit 1 or Orbit 2** | | | | |
| $\chi(\Gamma_{perm})$ | 2 | 0 | 0 | 0 |
| $\chi(\tilde{V})$ | 3 | -1 | -1 | -1 |
| $\chi(\Gamma)$ | 6 | 0 | 0 | 0 |
| **Atom in site 4(g)/4(h)** | | | | |
| $\chi(\Gamma_{perm})$ | 4 | 0 | 0 | 0 |
| $\chi(\tilde{V})$ | 3 | -1 | -1 | -1 |
| $\chi(\Gamma)$ | 12 | 0 | 0 | 0 |

The matrix representations of the symmetry elements $\{m_x|0½0\}$ and $\{m_y|½00\}$ are purely imaginary. The unitary matrix $U = \begin{pmatrix}1 & 0\\0 & -i\end{pmatrix}$ transforms all the matrix representations of $\tau_1$ to real matrices as shown in the third row of Table IV. When $k_z \neq 0$, the positions of a Mn$^{4+}$ cation on the 4(f) site are split into two orbits: $(0,1/2,z)$, $(1/2,0,z)$ and $(0,1/2,-z)$, $(1/2,0,-z)$. This is because the mirror in the *ab*-plane is not an element of $G_k$. For each orbit the decomposition of the magnetic representation $\Gamma$ is $\Gamma = 3\Gamma_1$. For the Mn$^{3+}$ and $R$ atoms in positions 4(g) and 4(h), respectively, a single orbit exists and the decomposition of the magnetic structure is $\Gamma = 6\Gamma_1$. In both cases (position 4(f) and 4(g)/4(h)), the number of basis vectors projected is equal to the number of spin degrees of freedom.

The predicted space group for the ferroelectric phases, $Pb2_1m$,[9] is qualitatively consistent with the magnetic structures presented here. We have constructed a toy model using the Shubnikov formalism to describe the Mn spin configurations within a single crystallographic unit cell. The pairs of spins parallel to $c$, {$Mn^{4+}(1),Mn^{4+}(2)$} and {$Mn^{4+}(3),Mn^{4+}(4)$}, are always aligned in FM fashion, thus the mirror planes at $z = 0$ and $z = 1/2$ possess additional time reversal and are denoted by $m'$. A $b$-glide plane at $x = 1/4$ relates the $Mn^{4+}(1)$ moment at $(0,0.5,0.25+\delta)$ to the $Mn^{4+}(3)$ moment at $(0.5,0,0.25+\delta)$, reversing the sign of the $m_y$ component parallel to the glide plane. The $Mn^{4+}(3)$ moment at $(0.5,0,0.25+\delta)$ and the $Mn^{4+}(1)$ moment at $(1,0.5,0.25+\delta)$ are related by a $b'$-glide plane at $x = 3/4$, which reverses the $m_x$ component. The pairs of spins {$Mn^{4+}(1),Mn^{4+}(4)$} and {$Mn^{4+}(2),Mn^{4+}(3)$} are related by $2_1'$ and $2_1$ screw axes, respectively. The arrangement of the $Mn^{3+}$ moments is also well described by this set of symmetry elements, which are uniquely consistent with the magnetic space group $P_{2a}b'2_1m'$, a subgroup of the paramagnetic space group $Pb2_1m$. The $P_{2a}b'2_1m'$ symmetry elements are shown schematically in Fig. 10.

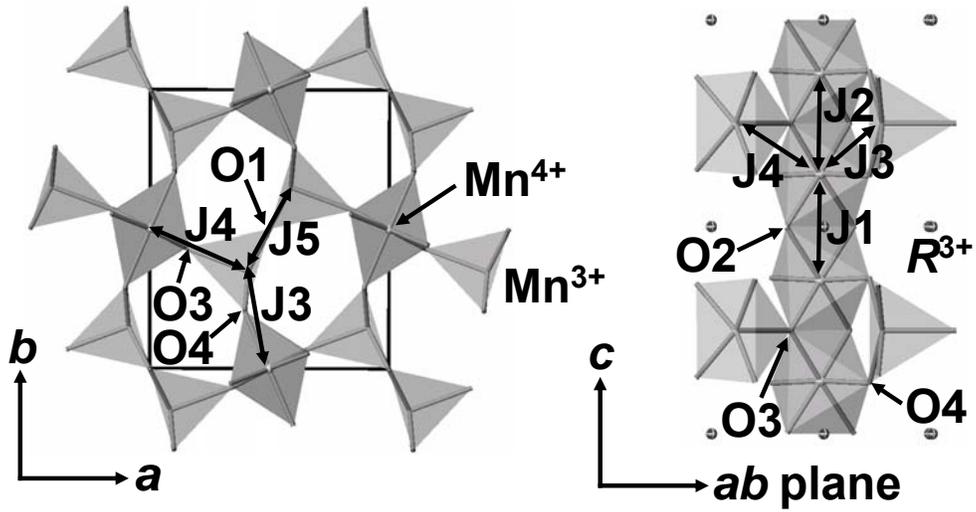

FIG. 1. Schematic crystal structure of $R$Mn$_2$O$_5$, showing magnetic exchange interactions referred to in the main text. Left: Mn$^{4+}$O$_6$ octahedra share corners with Mn$^{3+}$O$_5$ trigonal bipyramids in the *ab* plane; $R^{3+}$ cations are omitted for clarity. Right: Mn$^{4+}$O$_6$ octahedra share edges to form ribbons parallel to the *c* axis.

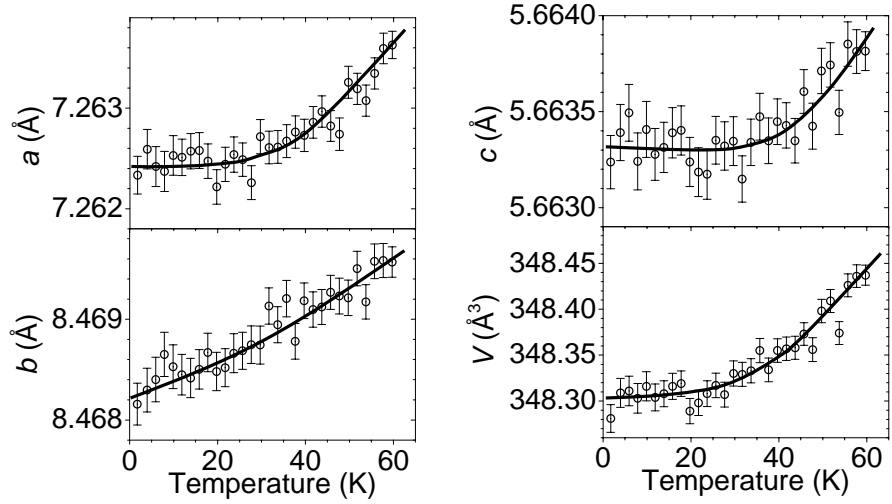

### a) HoMn$_2$O$_5$

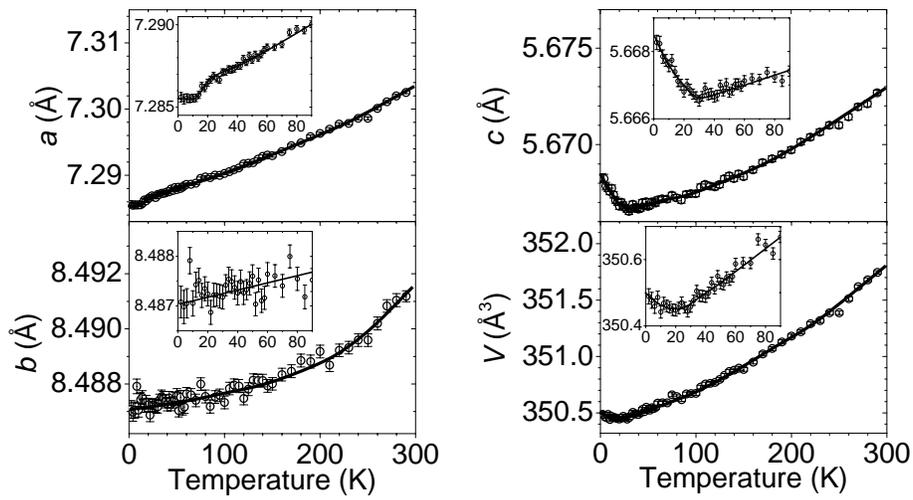

### b) DyMn$_2$O$_5$

FIG. 2. Lattice parameters and unit cell volumes of HoMn$_2$O$_5$ and DyMn$_2$O$_5$ as a function of temperature.

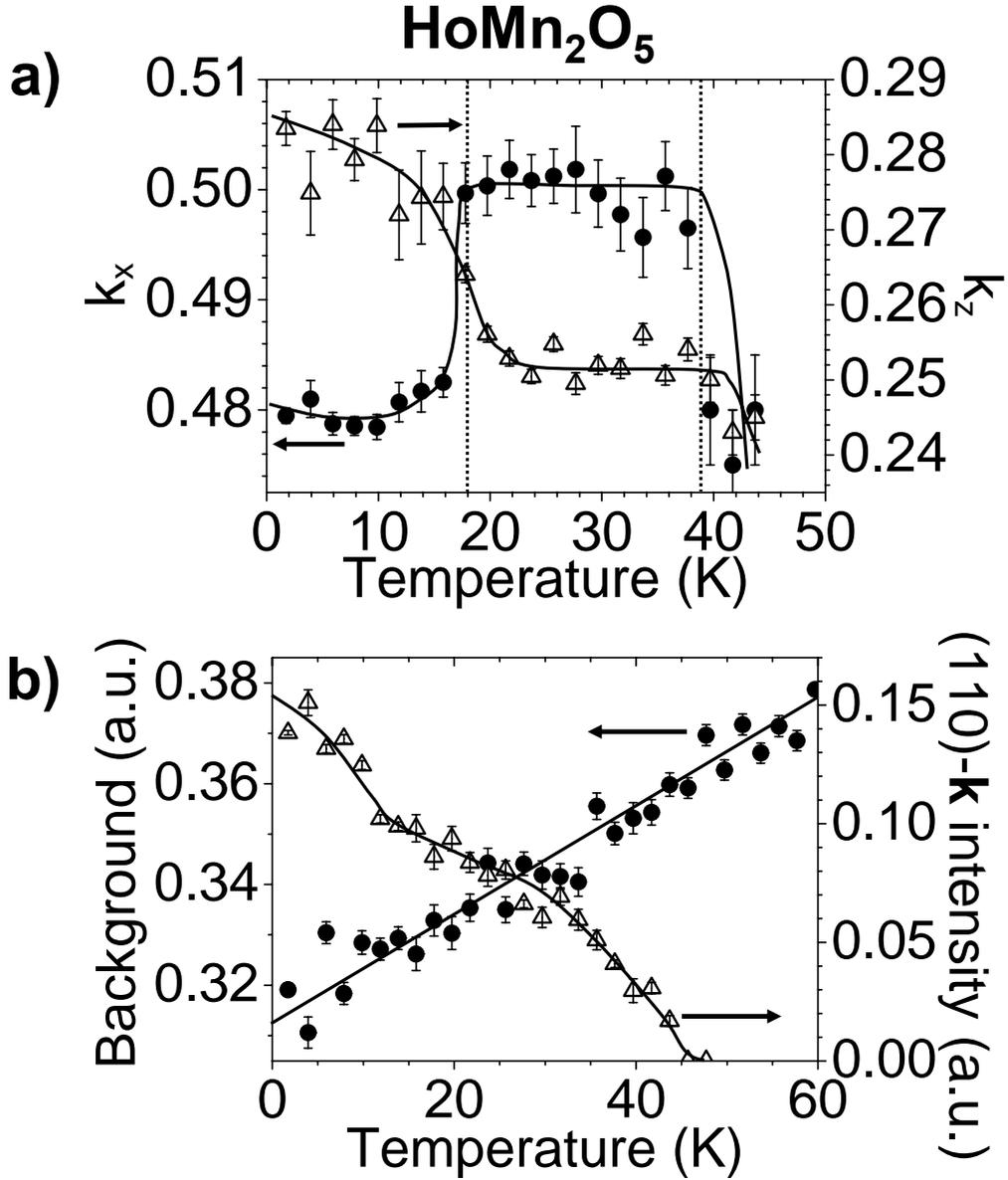

FIG. 3. (a) Magnetic propagation vector of HoMn$_2$O$_5$, **k** = ($k_x$,0,$k_z$), as a function of temperature; (b) Background and integrated intensity of (110)-**k** magnetic Bragg peak as a function of temperature for HoMn$_2$O$_5$.

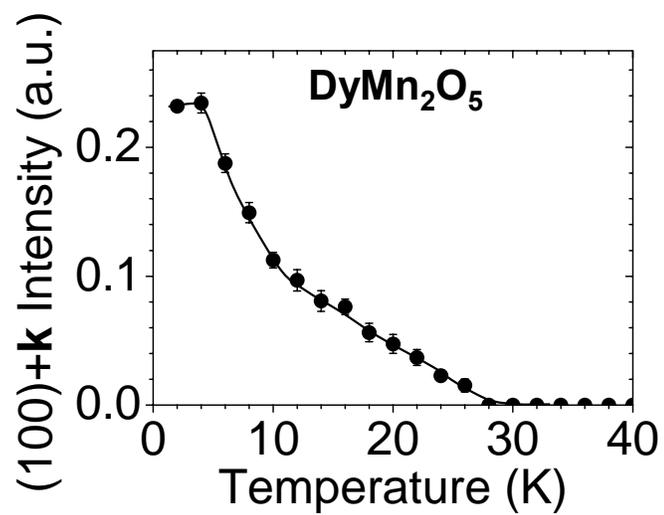

FIG. 4. Integrated intensity of (100)+**k** magnetic Bragg peak as a function of temperature for $DyMn_2O_5$.

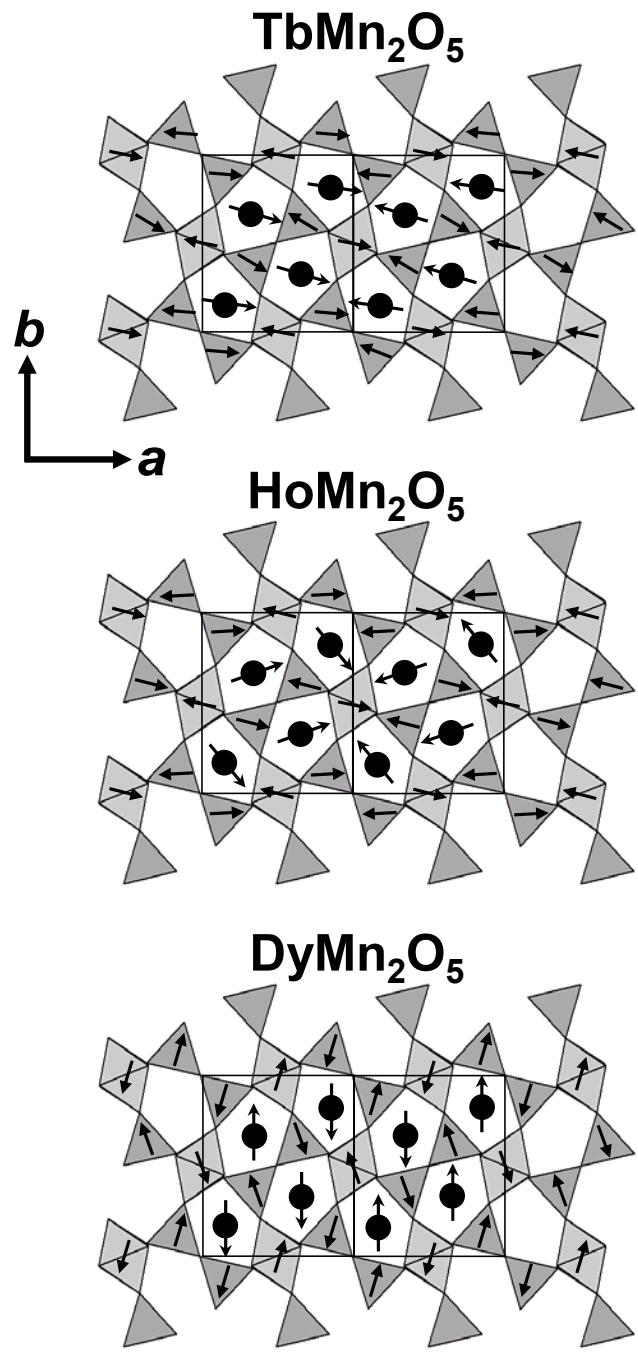

FIG. 5. Schematic representations of the magnetic structures of TbMn$_2$O$_5$, HoMn$_2$O$_5$ and DyMn$_2$O$_5$ in the $ab$ plane. The unit cells are doubled along $a$.

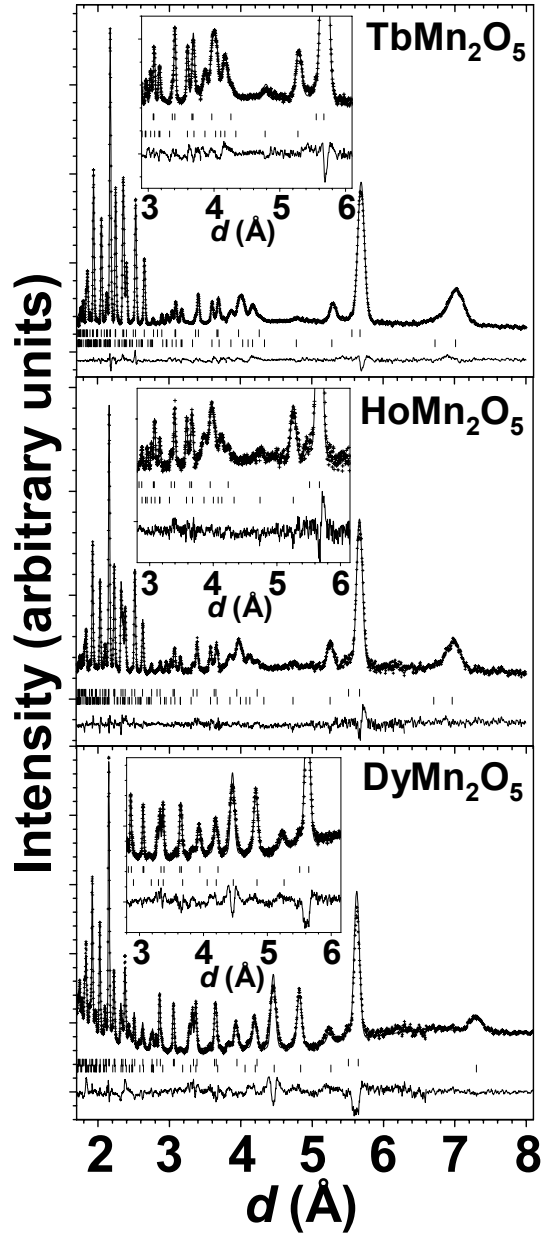

FIG. 6. Observed (crosses), calculated (solid line) and difference neutron diffraction profiles for TbMn$_2$O$_5$ at 27 K (top), HoMn$_2$O$_5$ at 26 K (middle), and DyMn$_2$O$_5$ at 2 K (bottom). The upper and lower rows of tick-marks correspond to reflection positions for the nuclear and magnetic structures, respectively. The data were collected from three detector banks situated at 18.0°, 35.0° and 63.6° and refined simultaneously. To produce the figure, data from different banks in adjacent $d$-spacing ranges were spliced at points of the profile where no Bragg peaks are present. The high quality of the fits to the weaker magnetic peaks is shown more clearly in the insets. Some weak unindexed peaks are apparent in the DyMn$_2$O$_5$ profile, as discussed in the main text.

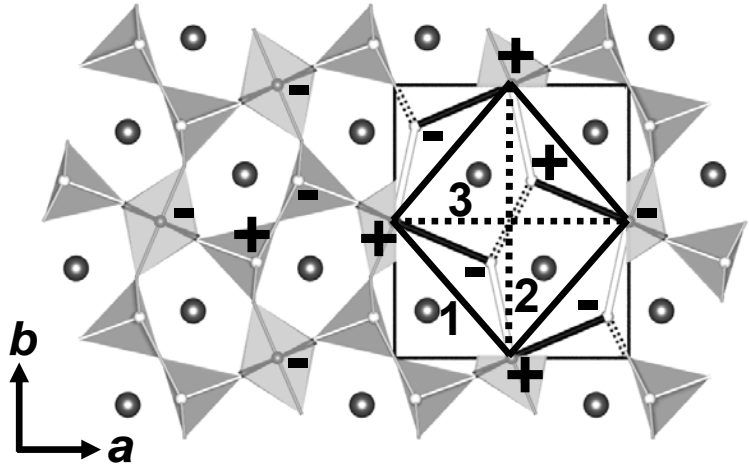

FIG. 7. Next-nearest-neighbor (NNN) magnetic exchange interactions in the *ab* plane. Spin directions are indicated by "+" and "-". Exchange interaction 3 is stronger than interaction 2, resulting in a square lattice of $Mn^{4+}$ with asymmetric NNN exchange and the stabilization of AFM zig-zag chains parallel to the *a* axis.

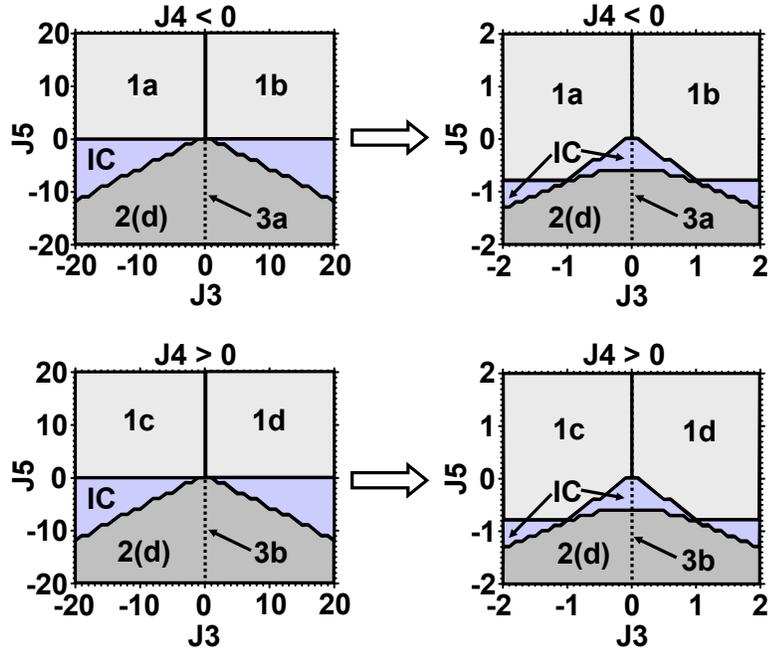

FIG. 8. Schematic magnetic phase diagrams calculated using ENERMAG, in space group *Pbam*. Exchange interactions J3 and J5 (see Fig. 1) are expressed in units of J4. Labels are as follows: structures 1a to 1d have **k** = (0,0) and spin configurations as listed in Table III; structures 2(d) have **k** = (0.5,0.5) and degenerate spin configurations; structures 3a and 3b have **k** = (0.5,0) and configurations as in Table III; structures IC are incommensurate.

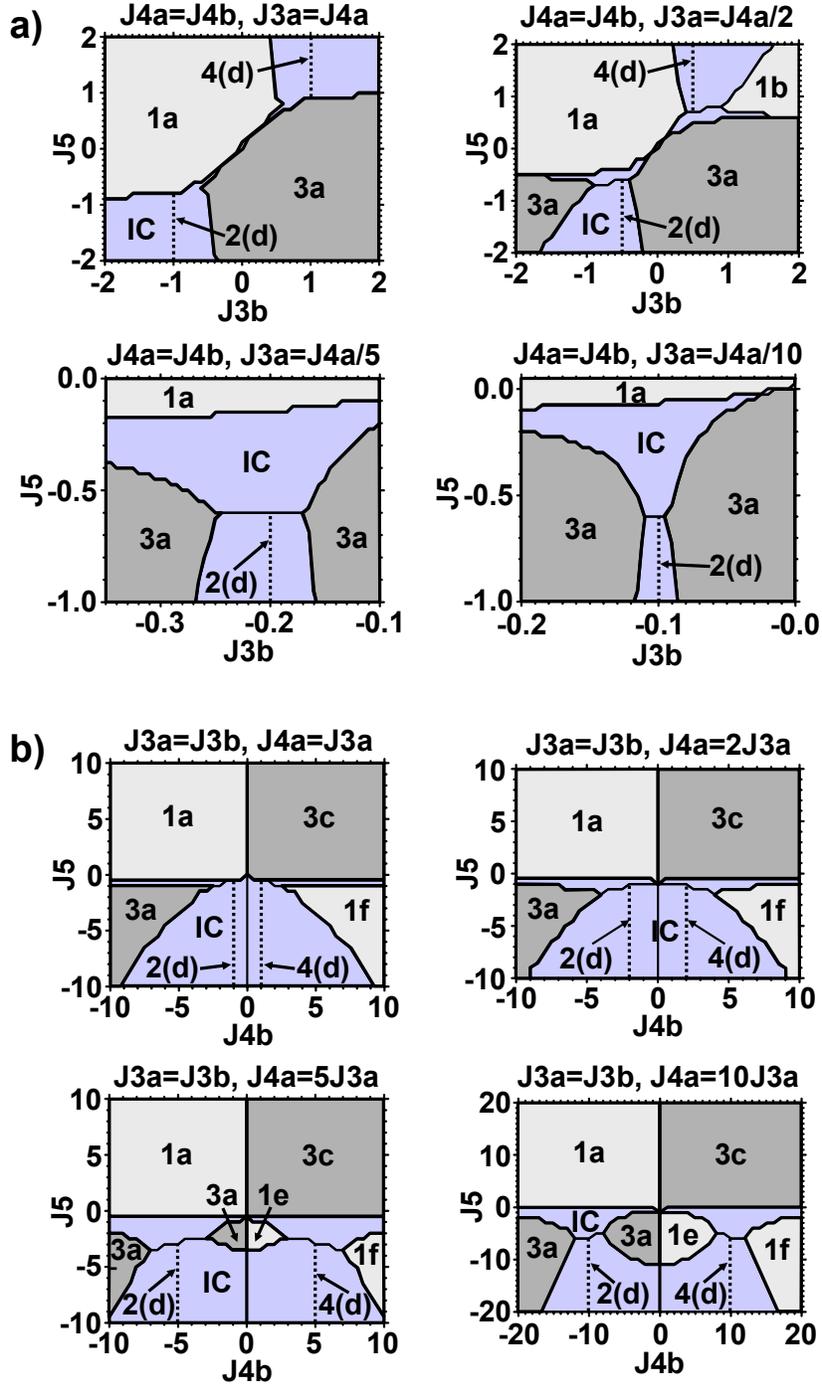

FIG. 9. Schematic magnetic phase diagrams calculated using ENERMAG, in space group $Pb2_1m$. Labels are as follows: structures 1a to 1f have $\mathbf{k} = (0,0)$ and spin configurations as listed in Table III; structures 2(d) have $\mathbf{k} = (0.5,0.5)$ and degenerate spin configurations; structures 3a and 3c have $\mathbf{k} = (0.5,0)$ and configurations as in Table III; structures 4(d) have $\mathbf{k} = (0,0.5)$ and degenerate spin configurations; structures IC are incommensurate. (a) The pair of exchange interactions J4a and J4b are equal and fixed. J3b and J5 are expressed in units of J4a (J4b). (b) The pair of exchange interactions J3a and J3b are equal and fixed. J4b and J5 are expressed in units of J3a (J3b).

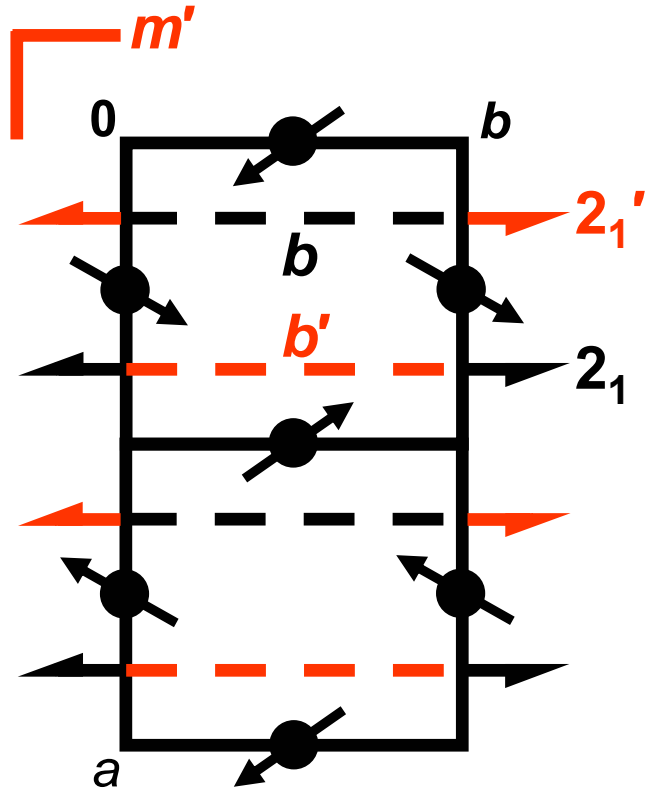

FIG. 10. Representation of the $P_{2a}b'2_1m'$ magnetic space group in the *ab* plane. The solid lines represent the unit cell with the *a* axis doubled. The $m'$, $2_1$ and $2_1'$ symmetry elements are situated at $z = 0$ and $z = \frac{1}{2}$. The $b$, $b'$, $2_1$ and $2_1'$ elements are at $x = \frac{1}{4}$ and $x = \frac{3}{4}$.